\documentclass[a4paper,USenglish,cleveref,autoref,thm-restate]{article}

\usepackage{amsthm,amssymb,amsmath}  
\usepackage{thmtools}
\usepackage{xspace}
\usepackage[utf8]{inputenc}
\usepackage{graphicx}
\usepackage[shortlabels]{enumitem}
\usepackage{hyperref}
\usepackage{cleveref}
\usepackage{fullpage}

\theoremstyle{plain}
\newtheorem{theorem}{Theorem}
\newtheorem{lemma}{Lemma}  
\newtheorem{corollary}{Corollary} 
\newtheorem{proposition}{Proposition}  
\newtheorem{fact}{Fact}
\newtheorem{claim}{Claim}
\newtheorem{observation}{Observation}

\theoremstyle{definition}
\newtheorem{definition}{Definition}

\newcommand{\THREESUM}{\textsc{3SUM}\xspace}
\newcommand{\THREESUMDS}{\textsc{3SUM Indexing}\xspace}
\newcommand{\THREESUMDSreport}{\textsc{3SUM Indexing w.~Reporting}\xspace}
\newcommand{\SSHIFT}{\textsc{Smallest Shift}\xspace}
\newcommand{\SSI}{\textsc{Shifted Set Intersection}\xspace}
\newcommand{\SSIreport}{\textsc{Shifted Set Intersection w.~Reporting}\xspace}
\newcommand{\SSIintervals}{\textsc{Gapped Set Intersection}\xspace}
\newcommand{\SSIintervalsrep}{\textsc{Gapped Set Intersection w.~Reporting}\xspace}
\newcommand{\SSIintervalsapprox}{\textsc{Approximate Gapped Set Intersection}\xspace}
\newcommand{\SSIintervalsapproxrep}{\textsc{Approximate Gapped Set Intersection w.~Reporting}\xspace}
\newcommand{\GI}{\textsc{Gapped String Indexing}\xspace}
\newcommand{\GIrep}{\textsc{Gapped String Indexing w.~Reporting}\xspace}
\newcommand{\JI}{\textsc{Jumbled Indexing}\xspace}
\newcommand{\JIrep}{\textsc{Jumbled Indexing w.~Reporting}\xspace}
\newcommand{\YES}{\textsf{YES}\xspace}
\newcommand{\NO}{\textsf{NO}\xspace}
\newcommand{\occ}{\mathrm{occ}}
\newcommand{\cO}{\mathcal{O}}
\newcommand{\ctO}{\mathcal{\tilde{O}}}
\def\dd{\mathinner{.\,.}}

\newcommand{\defDSproblem}[3]{
   \vspace{2mm}
 \noindent\fbox{
   \begin{minipage}{0.96\textwidth}
   \textsc{#1}\\
   {\bf{Preprocess:}} #2  \\
   {\bf{Query:}} #3
   \end{minipage}
   }
   \vspace{2mm}
}

\newenvironment{claimproof}[1]{\par\noindent\underline{Proof:}\space#1}{\hfill $\blacksquare$}

\title{Gapped String Indexing in Subquadratic Space and Sublinear Query Time}

\author{Philip Bille\\\texttt{phbi@dtu.dk} \and
Inge Li G{\o}rtz\\\texttt{inge@dtu.dk} \and
Moshe Lewenstein\\\texttt{moshe@cs.biu.ac.il} \and Solon P. Pissis\\\texttt{solon.pissis@cwi.nl}\and
Eva Rotenberg\\\texttt{erot@dtu.dk}\and
Teresa Anna Steiner\\\texttt{terst@dtu.dk}}

\begin{document}

\maketitle

\begin{abstract}
In \textsc{Gapped String Indexing}, the goal is to compactly represent a string $S$ of length $n$ such that for any query consisting of two strings $P_1$ and $P_2$, called \emph{patterns}, and an integer interval $[\alpha, \beta]$, called \emph{gap range}, we can quickly find occurrences of $P_1$ and $P_2$ in $S$ with distance in $[\alpha, \beta]$. \textsc{Gapped String Indexing} is a central problem in computational biology and text mining and has thus received significant research interest, including parameterized and heuristic approaches. Despite this interest, the best-known time-space trade-offs for  \textsc{Gapped String Indexing} are the straightforward $\mathcal{O}(n)$ space and $\mathcal{O}(n+ \mathrm{occ})$ query time or $\Omega(n^2)$ space and $\mathcal{\tilde{O}}(|P_1| + |P_2| + \mathrm{occ})$ query time.

We break through this barrier obtaining the first interesting trade-offs with  polynomially subquadratic space and polynomially sublinear query time. In particular, we show that, for every $0\leq \delta \leq 1$, there is a data structure for \textsc{Gapped String Indexing}\xspace with either $\mathcal{\tilde{O}}(n^{2-\delta/3})$ or $\mathcal{\tilde{O}}(n^{3-2\delta})$ space and $\mathcal{\tilde{O}}(|P_1| + |P_2| + n^{\delta}\cdot (\mathrm{occ}+1))$ query time, where $\mathrm{occ}$ is the number of reported occurrences.

As a new fundamental tool towards obtaining our main result, we introduce the \textsc{Shifted Set Intersection} problem: preprocess a collection of sets $S_1, \ldots, S_k$ of integers such that for any query consisting of three integers $i,j,s$, we can quickly output \textsf{YES} if and only if there exist $a \in S_i$ and $b \in S_j$ with $a+s = b$. 
We start by showing that the \textsc{Shifted Set Intersection} problem is equivalent to the indexing variant of 3SUM (\textsc{3SUM Indexing}\xspace) [Golovnev et al., STOC 2020]. We then give a data structure for \textsc{Shifted Set Intersection} with gaps, which entails a solution to the \textsc{Gapped String Indexing}\xspace problem. Furthermore, we enhance our data structure for deciding \textsc{Shifted Set Intersection}, so that we can support the reporting variant of the problem, i.e., outputting all certificates in the affirmative case. Via the obtained equivalence to \textsc{3SUM Indexing}\xspace, we thus give new improved data structures for the reporting variant of \textsc{3SUM Indexing}, and we show how this improves upon the state-of-the-art solution for \textsc{Jumbled Indexing} [Chan and Lewenstein, STOC 2015] for any alphabet of constant size $\sigma>5$.
\end{abstract}

\section{Introduction}\label{sec:intro}
The classic string indexing (or text indexing) problem~\cite{DBLP:books/cu/Gusfield1997,DBLP:books/daglib/0020103} is to preprocess a string $S$ into a compact data structure that supports efficient pattern matching queries; i.e., \emph{decide} if the pattern occurs or not in $S$ or \emph{report} the set of all positions in $S$ where an occurrence of the pattern starts. An important variant of practical interest is the \GI problem; the goal is to preprocess a string $S$ of length $n$ into a compact data structure, such that for any query consisting of two patterns $P_1$ and $P_2$, and a gap range $[\alpha, \beta]$, one can quickly find occurrences of $P_1$ and $P_2$ in $S$ with distance in $[\alpha, \beta]$.

Searching for patterns with gaps is of great importance in computational biology~\cite{BB1994,HPFB1999,FG2008,Myers1996,MM1993a,journals/jcb/NavarroR03,DBLP:journals/bmcbi/Pissis14}. In DNA sequences, a structured DNA motif consists of two smaller conserved sites (patterns $P_1$ and $P_2$) separated by a spacer, that is, a non-conserved spacer of mostly fixed or slightly variable length (gap range $[\alpha, \beta]$). Thus, by introducing an efficient data structure for \GI, one can preprocess a DNA sequence into a compact data structure that facilitates efficient subsequent searches to DNA motifs.

Searching for patterns with gaps appears also in the area of text mining~\cite{NLP,grammar,textmining,DBLP:conf/dasfaa/WillkommSB21}. Here, the task of finding co-occurrences of words is important, because they can indicate semantic proximity or idiomatic expressions in the text. A co-occurrence is a sequence of words (patterns $P_1,P_2,\ldots , P_k$) that occur in close proximity (gap range $[\alpha, \beta]$ or, most often, gap range $[0, \beta]$). By giving a data structure for \GI, one can pre-process large bodies of text into a compact data structure that facilitates efficient subsequent co-occurrence queries for the first non-trivial number of patterns, i.e., for $k=2$.

The algorithmic version of the problem, where a string and a \emph{single query} is given as input, is well-studied~\cite{conf/soda/BilleT10,BGVW2012,FG2008,journals/jcb/MorgantePVZ05,journals/jcb/NavarroR03,journals/cacm/Thompson68}. The indexing version, which is arguably more useful in real-world applications, is also much more challenging: Standard techniques yield an $\cO(n)$-space and $\cO(n+\occ)$-query time solution (see Appendix~\ref{sec:linear_space}) or an $\cO(n^3)$-space and $\cO(|P_1|+|P_2|+\occ)$-query time solution, where $\occ$ is the size of the output. A more involved approach yields $\ctO(n^2)$ space and $\ctO(|P_1|+|P_2|+\occ)$ query time (see Appendix~\ref{sec:quadratic_space}). 

The Combinatorial Pattern Matching community has been unable to improve the above-mentioned long-standing trade-offs. Precisely because of this, practitioners have typically been engineering algorithms or studying heuristic approaches~\cite{bader2016practical,BB1994,caceres2020fast,FG2008,HSSS2011,textmining,journals/jcb/MorgantePVZ05,Myers1996,journals/jcb/NavarroR03,DBLP:journals/bmcbi/Pissis14,DBLP:conf/dasfaa/WillkommSB21}. Theoreticians, on the other hand, have typically been solving restricted or parameterized variants~\cite{journals/ijfcs/PeterlongoAS08,bille2014substring,iliopoulos2009indexing,lewenstein2011indexing,bille2014string,cpm/KopelowitzK16,DBLP:journals/algorithmica/BilleGPS23,DBLP:conf/stoc/ColeGL04,DBLP:journals/tcs/LewensteinMRT14,DBLP:conf/esa/GawrychowskiLN14,DBLP:conf/stacs/LewensteinNV14}. Thus, breaking through the quadratic-space linear-time barrier for gapped string indexing is likely to have major consequences both in theory and in practice. This work is dedicated to answering the following question:
\begin{center}
    \emph{Is there a subquadratic-space and sublinear-query time solution for\\ \GI?}
\end{center}

\subsection{Results}
We assume throughout the standard word-RAM model of computation and answer the above question in the affirmative. Let us start by formally defining the \emph{existence variant} of \GI as follows: 

\defDSproblem{\GI}{A string $S$ of length $n$.}{Given two strings $P_1$ and $P_2$ and an integer interval $[\alpha, \beta]$, output \YES if and only if there exists a pair $(i,j)$ such that $P_1$ occurs at position $i$ of $S$, $P_2$ occurs at position $j\geq i$ of $S$ and $j-i \in [\alpha, \beta]$.}

Similarly, in the \emph{reporting variant} of \GI, a query answer is all pairs satisfying the above conditions.

\defDSproblem{\GIrep}{A string $S$ of length $n$.}{Given two strings $P_1$ and $P_2$ and an integer interval $[\alpha, \beta]$, output all pairs $(i,j)$ such that $P_1$ occurs at position $i$ of $S$, $P_2$ occurs at position $j\geq i$ of $S$ and $j-i \in [\alpha, \beta]$.}

\noindent Our main result is the following trade-offs for \GI with reporting:

\begin{theorem}\label{the:main}For every $0\leq \delta\leq 1$, there is a data structure for \GIrep with either:
\begin{description}
\item[(i)] $\ctO(n^{2-\delta/3})$ space and $\ctO(n^{\delta}\cdot (\occ+1) + |P_1| + |P_2|)$ query time; or 
\item[(ii)] $\ctO(n^{3-2\delta})$ space and $\ctO(n^{\delta}\cdot (\occ+1) + |P_1| + |P_2|)$ query time,
\end{description} 
where $\occ$ is the size of the output.
\end{theorem}
For the existence variant, the bounds above hold with $\occ=1$: we can terminate the querying algorithm as soon as the first witness is reported.
(Note that $n^{3-2\delta}$ is smaller than $n^{2-\delta/3}$ for $\delta>3/5$.)
Hence, we achieve the first polynomially \emph{subquadratic space} and polynomially \emph{sublinear query time} for \GI. 

Our main technical contribution is a data structure for \SSIintervalsrep, which is a generalization of a problem related to \THREESUMDS. We then show that our improved result for \GIrep follows via new connections to \SSIintervalsrep.

We show that the \SSIintervals and \THREESUMDS problems (and their reporting variants) are equivalent, but with an increase in universe size in the reduction to \THREESUMDS. Thus, as a result, we obtain a new data structure for the indexing variant of the 3SUM problem~\cite{DBLP:conf/stoc/GolovnevGHPV20}, which not only outputs an arbitrary certificate, but it outputs \emph{all certificates}. This is an interesting contribution on its own right and also has applications e.g., to the \JI problem~\cite{conf/stoc/ChanL15}. 

\subsection{Overview of Techniques and Paper Organization}

\begin{figure}[ht]
    \centering
    \includegraphics[width=0.9\textwidth]{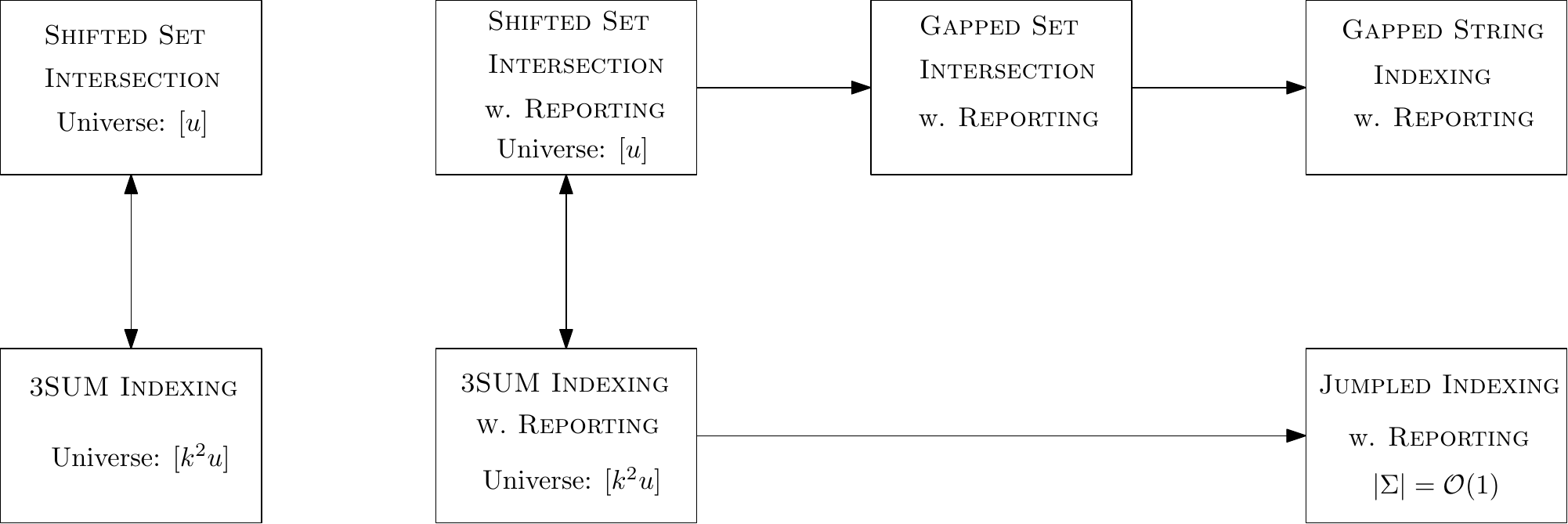}
    \caption{We show that \SSI and \THREESUMDS are equivalent, and our reduction also transfers to the reporting variants of the two problems. In fact, we solve a problem that is even harder than the \SSI problem; namely, the \SSIintervals problem, which leads to a solution to the \GI problem.}
    \label{fig:techniques}
\end{figure}

As a new fundamental tool towards obtaining our main result for \GI, we introduce the following problem, which we call \SSI (see~\cref{fig:techniques}).

\defDSproblem{\SSI}{A collection of $k$ sets $S_1, \ldots, S_k$ of total size $\sum_i |S_i| = N$ of integers from a universe $U = \{1,2, \dots, u\}$.}{Given $i,j,s$, output \YES\ if and only if there exist $a \in S_i$ and $b \in S_j$ such that $a+s = b$.}

We start by showing that the \SSI problem is equivalent to the indexing variant of \THREESUM. We formally define the \THREESUMDS problem next.

\defDSproblem{3SUM Indexing}{A set $A$ of $N$ integers from a universe $U = \{1,2, \dots, u\}$.}{Given $c\in U$ output \YES\ if and only if there exist $a, b \in A$ such that $a + b = c$.}

The following breakthrough result is known for \THREESUMDS; see also~\cite{DBLP:journals/corr/abs-1907-11206}.

\begin{theorem}[Golovnev~et~al.~\cite{DBLP:conf/stoc/GolovnevGHPV20}]\label{the:3SUMindexing}
For every $0\leq \delta \leq 1$, there is a data structure for \THREESUMDS with space $\ctO(N^{2-\delta})$ and query time $\ctO(N^{3\delta})$. In particular, this data structure returns a \emph{certificate}, i.e., for a query $c$, such that the answer is \YES, it can output a pair $a,b\in A$ such that $a+b=c$ in $\ctO(N^{3\delta})$ time.
\end{theorem}

In Section~\ref{sec:reductions}, we show a two-way linear-time reduction between \SSI and \THREESUMDS. In particular, this tells us that the two problems admit the same space-query time trade-offs. While the direction from \THREESUMDS to \SSI is immediate, the opposite direction requires some careful manipulation of the input collection based on the underlying universe. Furthermore, in the same section, we give a different trade-off for \SSI in the case where the set of possible distances is small, which is based on tabulating large input sets.

The main advantage of the \SSI formulation is that it gives extra flexibility which we can exploit to solve several generalizations of the problem. In particular, in Section~\ref{sec:reporting}, we show that we can augment any \SSI instance of $k$ sets and total size $N$ by $\cO(k\log N)$ extra sets, such that we can represent any subset of consecutive elements in a set of the original instance by at most $\cO(\log N)$ sets in the augmented instance. We use this to solve the reporting variant of the \SSI problem, called \SSIreport, where instead of deciding whether two elements of a given shift exist, we report all such elements. The idea is to first locate one output pair using the solution for the existence variant, then split the sets into elements smaller or bigger than the output element, and recurse accordingly.

\defDSproblem{\SSIreport}{A collection of $k$ sets $S_1, \ldots, S_k$ of total size $\sum_i |S_i| = N$ of integers from a universe $U = \{1,2, \dots, u\}$.}{Given $i,j,s$, output all pairs $(a,b)$ such that $a \in S_i$ and $b \in S_j$ and $a+s = b$.}

We show the following result.
    
\begin{restatable}{theorem}{reporting}\label{thm:reporting}
For every $0\leq \delta\leq 1$, there is a data structure for \SSIreport\ with $\ctO(N^{2-\delta/3})$ space and $\ctO(N^{\delta}\cdot (\occ+1))$ query time, where $\occ$ is the size of the output.
\end{restatable}    
    
As a consequence, we also give a reporting data structure for \THREESUMDS, which may be of independent interest. Let us first formalize the problem.

\defDSproblem{\THREESUMDSreport}{A set $A$ of $N$ integers from a universe $U = \{1,2, \dots, u\}$.}{Given $c\in U$ output all pairs $(a,b)\in A$ such that $a + b = c$.}

\begin{restatable}{corollary}{threesumreporting}\label{thm:3sum-reporting}
  For every $0\leq \delta\leq 1$, there is a data structure for \THREESUMDSreport\ with $\ctO(N^{2-\delta/3})$ space and $\ctO(N^{\delta}\cdot (\occ+1))$ query time, where $\occ$ is the size of the output.
\end{restatable}     

Furthermore, in Section~\ref{sec:intervals}, we show that we can augment any \SSI instance by $\cO(k\log u)$ sets to solve the more general problem of \SSIintervals, where instead of a \emph{single shift}, the query asks for an interval of allowed shifts.

\defDSproblem{\SSIintervals}{A collection of $k$ sets $S_1, \ldots, S_k$ of total size $\sum_i |S_i| = N$ of integers from a universe $U = \{1,2, \dots, u\}$.}{Given $i,j$ and an integer interval $[\alpha,\beta]$, output \YES\ if and only if there exist $a \in S_i$, $b \in S_j$ and $s\in [\alpha,\beta]$ such that $a+s = b$.}

We solve this problem by first considering an approximate variant of the problem, where the interval length is a power of two plus one, and we allow false positives in an interval of roughly twice the size. Then we carefully cover the query interval by $\cO(\log u)$ approximate queries to obtain Theorem~\ref{thm:intervals}.

\begin{restatable}{theorem}{intervals}\label{thm:intervals}
For every $0\leq \delta\leq 1$, there is a data structure for \SSIintervals\ with $\ctO(N^{2-\delta/3})$ space and $\ctO(N^{\delta})$ query time.
\end{restatable}  

We combine the reduction underlying Theorem~\ref{thm:intervals} with the reporting solution underlying Theorem~\ref{thm:reporting} to obtain a solution for the reporting variant of \SSIintervals.
        
\defDSproblem{\SSIintervalsrep}{A collection of $k$ sets $S_1, \ldots, S_k$ of total size $\sum_i |S_i| = N$ of integers from a universe $U = \{1,2, \dots, u\}$.}{Given $i,j$ and an integer interval $[\alpha,\beta]$, output all pairs $(a,b)$ such that $a \in S_i$, $b \in S_j$ and there is an $s\in [\alpha,\beta]$ such that $a+s = b$.}

\begin{restatable}{theorem}{intervalsrep}\label{thm:intervalsrep}
    For every $0\leq \delta \leq 1$, there is a data structure for \SSIintervalsrep with $\ctO(N^{2-\delta/3})$ space and $\ctO(N^{\delta}\cdot(\occ+1))$ query time, where $\occ$ is the size of the output.
\end{restatable} 

In Section~\ref{sec:GI}, we finally use all of these acquired tools to solve the \GI problem. In particular, we cover the suffix array of string $S$ in dyadic subintervals, which we then preprocess into the data structure from Theorem~\ref{thm:intervalsrep}. Note that the total size of these sets is $\cO(n\log n)$, where~$n$ is the length of $S$. Moreover, any interval of consecutive elements in the suffix array can be covered by $\cO(\log n)$ sets in the instance. Putting everything together we obtain our main result (Theorem~\ref{the:main}). 

In Section~\ref{sec:JI}, we show a reduction from \THREESUMDS to \JI, giving a data structure for \JI\ which improves on the state-of-the-art solution for constant alphabet sizes $\sigma>5$. We give new upper bounds for the existence and the reporting variants of \JI. In order to define \JI, we need the following notion of a \emph{histogram} of a string:
\begin{definition}
    Given a string $S$ over an alphabet $\Sigma$, a histogram $h$ of $S$ is a vector of dimension $|\Sigma|$ where every entry is the number of times the corresponding letter occurs in $S$.  
\end{definition}
\defDSproblem{\JIrep}{A string $S$ of length $n$ over an alphabet $\Sigma$.}{For any pattern histogram $P\in (\mathbb{N}_0)^{|\Sigma|}$, return all substrings of $S$ whose histogram is $P$.}

We call a substring of $S$ whose histogram is $P$ an \emph{occurrence of $P$ in $S$}.
In the \emph{existence variant} of the \JI problem, the query answer is \YES or \NO depending on whether there is at least one occurrence of $P$ in $S$. We show the following result:

\begin{restatable}{corollary}{corjumbled}\label{cor:jumbled}
Given a string $S$ of length $n$ over an alphabet of size $\sigma=\cO(1)$, for every $0\leq \delta \leq 1$, there is a data structure for $\JIrep$ with $\ctO(n^{2-\delta})$ space and $\ctO(n^{3\delta}\cdot(\occ+1))$ query time, where $\occ$ is the size of the output. The bounds are $\ctO(n^{2-\delta})$ space and $\ctO(n^{3\delta})$ query time for the existence variant.
\end{restatable}

The best-known previous bounds for the existence variant of \JI\ use $\cO(n^{2-\delta})$ space and $\cO(m^{(2\sigma-1)\delta})$ query time~\cite{journals/algorithmica/KociumakaRR17}, where $m$ is the norm of the queried pattern (i.e., the sum of histogram entries), or $\ctO(n^{2-\delta})$ space and $\ctO(n^{\delta(\sigma+1)/2})$ query time~\cite{conf/stoc/ChanL15}. Interestingly, the reporting variant of \JI\ is, in general, significantly harder than the existence variant: There is a recent (unconditional) lower bound stating that any data structure for $\JIrep$ with $\cO(n^{0.5-o(1)}+\occ)$ query time needs $\Omega(n^{2-o(1)})$ space, and this holds even for a binary alphabet~\cite{conf/soda/AfshaniDKN20}. By our reduction, this in particular implies that a data structure with $\ctO(N^{2-\delta})$ space and $\ctO(N^{3\delta}+\occ)$ query time is not possible for \THREESUMDSreport. For a more complete overview, see Section~\ref{sec:JI}. We conclude this paper in Section~\ref{sec:con} with some future proposals. 

In Appendix~\ref{sec:smallest_shift}, we show a better trade-off for the related \SSHIFT problem. 

\defDSproblem{Smallest Shift}{A collection of $k$ sets $S_1, \ldots, S_k$ of total size $\sum_i |S_i| = N$ of integers from a universe $U = \{1,2, \ldots, u\}$.}{Given $i,j$, output 
the smallest $s$ such that there exists $a\in S_i$ and $b\in S_j$ with $a+s=b$.}
\begin{proposition}
There is a data structure for \SSHIFT with $\cO(N)$ space and $\ctO(\sqrt{N})$ query time. Moreover, the data structure can be constructed in $\cO(N \sqrt{N})$ time.
\end{proposition}

\SSIintervals reduces to \SSHIFT in the special case where we only allow query intervals of the form $[0,\beta]$. Together with our other results, this reduction yields a $\ctO(N)$ space and $\ctO(|P_1|+|P_2|+\sqrt{N}\cdot(\occ+1))$ query time trade-off for \GI if the query intervals are restricted to $[0,\beta]$. However, let us remark that 
for this restricted version of the problem, a $\ctO(N)$ space and $\ctO(|P_1|+|P_2|+\sqrt{N\cdot(\occ+1)}+\occ)$ query time trade-off already follows from~\cite{DBLP:journals/algorithmica/BilleGPS23}.

\subparagraph*{Subsequent work.}~Since the first publication of this paper~\cite{bille2022gapped}, Aronov~et~al.~\cite{sosa2024} proposed a combination of range searching and the Fiat-Naor inversion scheme~\cite{DBLP:journals/siamcomp/FiatN99} (which was has already been used to prove Theorem~\ref{the:3SUMindexing}), recovering our main result (Theorem~\ref{the:main}) as a corollary.

\section{\THREESUMDS is Equivalent to \SSI}\label{sec:reductions}

Here we show a two-way linear-time reduction between \THREESUMDS and \SSI.
We thus obtain the same space-time bounds for the two problems.

\subsection{From \THREESUMDS to \SSI}

\begin{fact}
Any instance of \THREESUMDS of input size $N$ and universe size $u$ can be reduced to a \SSI  instance of total size $\cO(N)$ with universe size $\cO(u)$ in time $\cO(N)$.
\end{fact}
\begin{proof}
We denote the \THREESUMDS instance by $A = \{a_1, a_2, \ldots, a_N\}$ (the input set).
We construct the \SSI instance: $S_1 = A$ and $S_2 = \{-a_1, -a_2,\dots,-a_N\}$ in $\cO(N)$ time.
For any \THREESUMDS query $c$, we construct the \SSI query $(S_2, S_1, c)$\footnote{We may write $(S_i, S_j, c)$ for a \SSI query instead of $(i, j, c)$ for clarity.} in $\cO(1)$ time, and observe that: $a_i + a_j = c \iff c - a_i = a_j$. Thus the answer to the \THREESUMDS query is \YES\ if and only if the answer to the \SSI query is \YES. 
\end{proof}

\subsection{From \SSI to \THREESUMDS}

\begin{theorem}
Any instance of \SSI with $\sum_{i=1}^k |S_i|=N$ and universe size $u$ can be reduced to a \THREESUMDS instance of input size $\cO(N)$ and universe size $\cO(k^2u)$ in time $\cO(N)$.
\end{theorem}

\begin{proof}
Let us start by considering an alternative yet equivalent formulation of \THREESUMDS. Preprocess two sets $A$ and $B$ from a universe $U'=\{1,2,\ldots,u'\}$ to answer queries of the following form: Given an element $c\in U'$, output \YES\ if and only if there exist $(a, b) \in A\times B$ such that $a + b = c$. To see why it is equivalent, notice that the formulation with one set reduces trivially to this formulation by setting $A=B$. For the other direction, given $A$ and $B$ from universe $U'$, define $A'=A\cup \{b+2u'~|~b\in B\}$. We verify that querying $c+2u'$ on $A'$ for any $c\in [2,\dots,2u']$ in the \THREESUMDS formulation with one set is equivalent to querying $c$ on $A$ and $B$ in the \THREESUMDS formulation with two sets. We now reduce \SSI to the  \THREESUMDS formulation with two sets.

We denote the \SSI instance by $S_1,\dots,S_k$ (the input collection over universe $U=\{1,2,\dots,u\}$) and $\sum_{i\in[k]}|S_i|$ by $N$.
We construct the following instance of the above \THREESUMDS reformulation in $\cO(N+k)=\cO(N)$ time:
    \begin{align*}
        A=\{e+j\cdot (k+1)\cdot 2u~|~e\in S_j,~1\leq j\leq k\},~B=\{-e+i\cdot 2u~|~e\in S_i, ~1\leq i\leq k\}.
    \end{align*}
The \THREESUMDS instance has input size $2N=\Theta(N)$ and universe size $U'=\cO(k^2\cdot u)$.

Let us denote a query of \SSI by $Q(S_i,S_j,s)$. 
Now, we construct the \THREESUMDS query $Q_{\text{3SUM}}(s+(j\cdot(k+1)+i)\cdot 2u)$ in $\cO(1)$ time.
The following claim concludes the proof.

\begin{claim}
$Q(S_i,S_j,s)$ outputs \YES\ if and only if $Q_{\text{3SUM}}(s+(j\cdot(k+1)+i)\cdot 2u)$ outputs \YES.
\end{claim}
\begin{claimproof}
$\mathbf{(\Rightarrow)}$: Let $e_1\in S_i$, $e_2\in S_j$ such that $e_1+s=e_2$. Then $a=e_2+j\cdot (k+1)\cdot 2u \in A$, $b=-e_1+i\cdot 2u \in B$, and $a+b=s+(j\cdot(k+1)+i)\cdot 2u$. Thus $Q_{\text{3SUM}}(s+(j\cdot(k+1)+i)\cdot 2u)$ outputs \YES.

$\mathbf{(\Leftarrow)}$: Assume there is $a\in A$ and $b\in B$ such that $a+b=s+(j\cdot(k+1)+i)\cdot 2u$. By definition of $A$ and $B$, there exist  $i'$, $j'$, and $e_2\in S_{j'}$, $e_1\in S_{i'}$ such that $a=e_2+j'\cdot(k+1)\cdot 2u$ and $b=-e_1+i'\cdot 2u$, thus $s+(j\cdot(k+1)+i)\cdot 2u=(e_2-e_1)+(j'\cdot(k+1)+i')\cdot 2u$. Since $-u<s<u$ and $-u<(e_2-e_1)<u$, we have that $j\cdot(k+1)+i=j'\cdot(k+1)+i'$ and $e_2-e_1=s$. Since $i\leq k$ and $i'\leq k$, we have $j=j'$ and $i=i'$. 
 Thus $Q(S_i,S_j,s)$ outputs \YES.   
\end{claimproof}
\end{proof}

By employing~\cref{the:3SUMindexing} we obtain the following corollary:

\begin{corollary}\label{cor:SSI}
    For every $0\leq \delta\leq 1$, there is a data structure for \SSI\ with space $\ctO(N^{2-\delta})$ and query time $\ctO(N^{3\delta})$.
\end{corollary}

We also show how we can obtain a different trade-off in the case where the number of possible shifts is bounded by some $\Delta$ (this gives us a better trade-off for some $\delta$ in the application of \GI). 
In this case, call all sets of size at most $N^{\delta}$ \emph{small}, and other sets \emph{large}. Note that there are at most $N^{1-\delta}$ large sets. For every pair of large sets and any possible shift, we precompute the answer, using space $\cO(\Delta N^{2-2\delta})$. Additionally, we store a dictionary on every set using space $\cO(N)$ (using e.g., perfect hashing~\cite{FKS1984}). For a query $Q(S_i,S_j, s)$, if both sets $S_i$ and $S_j$ are large, we look up the precomputed answer. If one set is small, wlog $S_i$, we check if $a+s\in S_j$ for every $a\in S_i$ using the dictionary. Note that in particular, $\Delta<u$. This gives the following lemma:

\begin{lemma}\label{lem:diftrade-off}
For every $0\leq \delta\leq 1$, there is a data structure for \SSI\ with space $\cO(u\cdot N^{2-2\delta})$ and query time $\cO(N^{\delta})$.
\end{lemma}

\section{Reporting: \THREESUMDS and \SSI}\label{sec:reporting}

In this section, we explain how we can answer reporting queries for both \SSI and \THREESUMDS, as defined in Section~\ref{sec:intro}. The following fact is trivial.

\begin{fact}
There is a data structure for \SSIreport\ with $\cO(N)$ space and $\cO(|S_i|+|S_j|)$ query time.
\end{fact}

The other extreme trade-off is also straightforward.

\begin{lemma}
There is a data structure for \SSIreport\ with $\cO(N^2)$ space and $\cO(\occ)$ query time, where $\occ$ is the size of the output.
\end{lemma}
\begin{proof}
    For every existing shift $s$, we save all the pairs of sets $(A, B)$ for which $Q(A,B, s) = $\YES\ using perfect hashing~\cite{FKS1984}. For every such pair, we save all pairs $(a,b) \in A\times B$, such that $a+s = b$, in a list. There are $\cO(N^2)$ pairs $(a,b)$ in total. The space is thus $\cO(N^2)$. 
\end{proof}

We next prove the main result (Theorem~\ref{thm:reporting}) of this section.

\reporting*
\threesumreporting*

\begin{proof}[Proof of Theorem~\ref{thm:reporting}]
  The main idea behind the proof is the following:
  We use the fact that the data structure for \THREESUMDS from Theorem~\ref{the:3SUMindexing}~returns a \emph{certificate}, i.e., for a query $c\in U$ such that the answer to the query is \YES\ it can additionally output a pair $(a,b)$ satisfying $a+b=c$. By our reduction from \SSI, when we query for $S_i$, $S_j$ and $s$ we can return $a\in S_i$ and $b\in S_j$ such that $a+s=b$. We then conceptually split $S_i$ and $S_j$ into two sets each, those elements in $S_i$ which are smaller than $a$, those which are bigger, and similarly for $S_j$ and $b$. We then want to use the fact that if $a'+s=b'$ and $a'<a$, then $b'<b$ and vice versa, to recurse on the smaller subsets. However, we cannot afford to preprocess all subsets of any set $S_i$ into the \SSI\ data structure. To solve this issue,
  we partition each set into the subsets which correspond to the dyadic intervals on the rank of the elements in sorted order, and preprocess them into our \SSI\ data structure. 
  
  In detail, our data structure is defined as follows: Let $S_1,\dots, S_k$ be the input to \SSIreport. We build the data structure for \SSI\ from Corollary~\ref{cor:SSI} containing, additionally to $S_1,\dots, S_k$, the following subsets:
        \item Let $S=\{s_1,s_2,  \ldots, s_m\}$ be a set in our \SSIreport\ instance, where $s_1 < s_2< \cdots < s_m$.

We partition $S$ into subsets which correspond to the dyadic intervals on the rank. That is, for $j=0,\dots,\lfloor \log m\rfloor$, we partition $S$ into $\{s_{i+1},s_{i+2},\dots,s_{i+2^j}\}$, for all $i=\kappa\cdot 2^j$ and $0\leq \kappa\leq \lfloor m/2^j\rfloor -1$. We call these sets the \emph{dyadic subsets} of $S$. Note that for a fixed $j$, the union of these sets is a subset of $S$ and thus has size at most $m$. Hence, the total size of all dyadic subsets is $\cO(m\log m)$.
\begin{observation}\label{obs:dyadic}
Any subset of $S$ of the form $\{x\in S~|~a\leq x \leq b\}$ is the union of at most $2\log |S|$ dyadic subsets of $S$.
\end{observation}

The data structure for \SSIreport\ consists of the \SSI\ data structure on all sets $S_1,\dots, S_k$ and their dyadic subsets. Further, for each dyadic subset, we store its minimum and its maximum element as auxiliary information. 
The total size of all dyadic subsets is $\cO(N\log N)$. The space is thus $\ctO(N^{2-\delta/3})$.

To perform a query on $S_i, S_j, s$, we first perform a query on the \SSI\ data structure for the same inputs. If it returns \NO, we are done. If it returns \YES, let $a$ and $b$ be the certificate pair such that $a+s=b$. Define $A_1=\{a'\in S_i~|~a'<a\}$ and $A_2=\{a'\in S_i~|~a'>a\}$, and similarly $B_1=\{b'\in S_j~|~b'<b\}$ and $B_2=\{b'\in S_j~|~b'>b\}$. Then any additional solution pair $(a',b')$ has to either satisfy $a'\in A_1$ and $b'\in B_1$ or $a'\in A_2$ and $b'\in B_2$.
Further, by Observation~\ref{obs:dyadic}, we can decompose $A_1$, $A_2$, $B_1$, and $B_2$ into $\cO(\log N)$ dyadic subsets in $\cO(\log N)$ time. Call these decompositions $\mathcal{A}_1$, $\mathcal{A}_2$, $\mathcal{B}_1$, $\mathcal{B}_2$. We could now recurse on any $(A,B)$ where $A\in \mathcal{A}_1$ and $B\in \mathcal{B}_1$ or $A\in \mathcal{A}_2$ and $B\in \mathcal{B}_2$, dividing into $\cO(\log^2 N)$ subproblems. Since every time before we recurse we obtain a new certificate, this brings the query time to $\ctO(N^{\delta}\cdot(\occ+1))$, which proves the theorem. We nevertheless show next (Lemma~\ref{lem:matchpairs}) how we can reduce the recursion pairs to $\cO(\log N)$.

  Fix $i\in \{1,2\}$. For $A'\in \mathcal{A}_i$, let  $a_{\min}$ be its minimum element and $a_{\max}$ its maximum element. For $B'\in \mathcal{B}_i$ define $b_{\min}$ and $b_{\max}$ analogously. We say $A'$ and $B'$ \emph{match} if $[a_{\min}+s,a_{\max}+s]$ and $[b_{\min},b_{\max}]$ have a non-empty intersection.
  
 \begin{lemma}\label{lem:matchpairs} There are $\cO(\log N)$ matching pairs $(A',B')$, $A'\in \mathcal{A}_i$, $B'\in \mathcal{B}_i$, and $i\in \{1,2\}$.\end{lemma}
        \begin{proof} For a fixed $A'\in \mathcal{A}_i$, consider all $B'$ such that $A'$ matches $B'\in \mathcal{B}_i$. Since the intervals $[b_{\min},b_{\max}]$ are disjoint, all except at most two $B'$ that match $A'$ have the property that the corresponding interval $[b_{\min},b_{\max}]$ is fully contained in $[a_{\min}+s,a_{\max}+s]$. Hence, those $B'$ match \emph{only} $A'$. Let $\mathcal{A}_i=\{A_i^1,\dots, A_i^{k_1}\}$ and $|\mathcal{B}_i|=k_2$ with $k_1, k_2=\cO(\log N)$. The number of matching pairs is given by
       \[\sum_{j=1}^{k_1}(\textnormal{Number of }B'\textnormal{ matching }A_i^j)
            \leq 2k_1+\sum_{j=1}^{k_1}(\textnormal{Number of }B'\textnormal{ matching only }A_i^j)\leq 2k_1+k_2.
\]
 The last inequality follows since the sets $\{B' \textnormal{~matching only~} A_i^j\}$ form disjoint subsets of~$\mathcal{B}_i$.
        \end{proof}
   For one $A'$, we can identify all matching $B'$ in $\cO(\log N)$ time using the precomputed information (i.e.,  searching for the predecessor of $a_{\min}+s$ in the minima of $B_i$ and the successor of $a_{\max}+s$ in the maxima of $B_i$). In total all matching pairs can be identified in time $\cO(\log^2N)$.
 Thus, instead of recursing on all pairs $(A', B')$, where $A' \in \mathcal{A}_i$ and $B' \in \mathcal{B}_i$, we can recurse only on the matching pairs, which saves a $\log N$ factor in the query time.
   \end{proof}

\section{\SSIintervals}\label{sec:intervals}
Here we show how to solve the more general problem of \SSIintervals, where we allow any shift within a given interval. Recall that the problem is defined in Section~\ref{sec:intro}.

The main idea is to use the solution from Corollary~\ref{cor:SSI} on $\cO(\log u)$ carefully constructed \SSI instances. To do this,  we first show how we can approximately answer \SSIintervals queries for query intervals whose length is a power of two plus one (approximately in the sense that we allow false positives within a larger interval), then use $\cO(\log u)$ such queries to ``cover'' $[\alpha,\beta]$. 

We formally define \emph{approximate} \SSIintervals. 

\defDSproblem{\SSIintervalsapprox}{A collection of $k$ sets $S_1, \ldots, S_k$ of total size $\sum_i |S_i| = N$ of integers from a universe $U = \{1,2, \dots, u\}$ and a \emph{level} $l$ with $1\leq l\leq \log u$.}{Given $i,j$ and a \emph{center distance} $d=\kappa\cdot 2^l$, for some positive integer $\kappa$, output \YES\ or \NO\ such that: 
\begin{itemize}
    \item The output is \YES if there exists a pair $a\in S_i$, $b\in S_j$ such that $b-a\in[d-2^{l-1},d+2^{l-1}]$;
    \item The output is \NO if there exists no pair $a\in S_i$, $b\in S_j$ such that $b-a\in(d-2^l,d+2^l)$.
\end{itemize}}

Note that by the definition of the \SSIintervalsapprox\ problem, if for a query $i, j$ and $d=\kappa\cdot 2^l$ there exists a pair  a pair $a\in S_i$, $b\in S_j$ such that $b-a\in(d-2^l,d+2^l)$, but no pair $a\in S_i$, $b\in S_j$ such that $b-a\in[d-2^{l-1},d+2^{l-1}]$, we may answer either \YES\ or \NO.  

    \begin{lemma}\label{lem:approxinterval}
    Assume there is a data structure for \SSI with $s$ space and $t$ query time. Then there is a data structure for \SSIintervalsapprox with ${\cO}(s)$ space and ${\cO}(t)$ query time.
    \end{lemma}
\begin{proof}
    For any set $S_i$ in the collection, define $S'_i=\{\lfloor a/2^{l-1} \rfloor, a\in S_i\}$
    and build the assumed data structure for \SSI on sets $S'_1,\dots,S'_k$. To answer a query $(i,j,d=\kappa\cdot 2^l)$ on \SSIintervalsapprox, make the queries $(i,j,2\kappa-1)$, $(i,j,2\kappa)$ and $(i,j,2\kappa+1)$ on the \SSI instance. Return \YES\ if and only if one of the queries to the \SSI instance returns \YES. 

    To show correctness, first notice that if we return \YES, then there exist $a\in S_i$ and $b\in S_j$ satisfying  $2\kappa-1\leq \lfloor b/2^{l-1}\rfloor - \lfloor a/2^{l-1}\rfloor \leq 2\kappa+1$.
    This implies $2\kappa-2\leq  \lfloor b/2^{l-1}\rfloor - \lfloor a/2^{l-1}\rfloor-1 < \lfloor b/2^{l-1}\rfloor - a/2^{l-1} \leq b/2^{l-1}- a/2^{l-1}$
    and 
    $b/2^{l-1}- a/2^{l-1}\leq b/2^{l-1} - \lfloor a/2^{l-1}\rfloor < \lfloor b/2^{l-1}\rfloor - \lfloor a/2^{l-1}\rfloor +1\leq 2\kappa+2$.
    Thus, $b-a\in (\kappa\cdot 2^l-2^l,\kappa\cdot 2^l+2^l)=(d-2^l, d+2^l)$.

    Next, assume there is a pair $a\in S_i$ and $b\in S_j$ such that $b-a\in[d-2^{l-1},d+2^{l-1}]=[\kappa\cdot 2^l-2^{l-1},\kappa\cdot 2^l+2^{l-1}]$. Then clearly $b/2^{l-1}- a/2^{l-1}\in [2\kappa-1,2\kappa+1]$. Similar to above, we have 
    $\lfloor b/2^{l-1}\rfloor - \lfloor a/2^{l-1}\rfloor \leq b/2^{l-1} - \lfloor a/2^{l-1}\rfloor<b/2^{l-1}-a/2^{l-1}+1\leq 2\kappa+2$
    and 
    $2\kappa-2\leq b/2^{l-1}- a/2^{l-1}-1<\lfloor b/2^{l-1}\rfloor- a/2^{l-1}\leq \lfloor b/2^{l-1}\rfloor - \lfloor a/2^{l-1}\rfloor $.
    Thus, $\lfloor b/2^{l-1}\rfloor - \lfloor a/2^{l-1}\rfloor\in (2\kappa-2, 2\kappa+2)$ and, because $\lfloor b/2^{l-1}\rfloor - \lfloor a/2^{l-1}\rfloor$ is an integer, $\lfloor b/2^{l-1}\rfloor - \lfloor a/2^{l-1}\rfloor\in[2\kappa-1,2\kappa+1]$. Thus we answer \YES\ in this case.

\end{proof}

\begin{corollary}
        For every $0\leq \delta \leq 1$, there is a data structure for \SSIintervalsapprox with $\ctO(N^{2-\delta/3})$ space and $\ctO(N^{\delta})$ query time.
    \end{corollary}
We now show how to reduce \SSIintervals to $\cO(\log u)$ \SSIintervalsapprox instances, giving the following theorem:

\intervals*

 \begin{proof} Given $S_1,\dots, S_k$, we build a \SSI data structure and an \SSIintervalsapprox data structure for $S_1,\dots, S_k$ and every level $l$ with $1\leq l\leq \log u$. Assume we want to answer a query for $S_i,S_j$ and $[\alpha,\beta]$. We show how to answer the query using $\cO(\log(\beta-\alpha))$ \SSIintervalsapprox queries.
We query \SSIintervalsapprox for $i,j$ and different choices of $l$ and $d$. We call a query $i,j$ and $d$ to \SSIintervalsapprox of level $l$  a \emph{$2^l$-approximate query centered at $d$}. We say the query \emph{covers} the interval $[d-2^{l-1},d+2^{l-1}]$. We call all elements in interval $(d-2^l,d+2^l)$ \emph{uncertain}. Now, for the interval $[\alpha,\beta]$, we want to find $\cO(\log (\beta-\alpha))=\cO(\log u)$ queries which together \emph{cover} $[\alpha,\beta]$ such that there are no uncertain elements outside of $[\alpha,\beta]$. In detail:
      
We show how to cover a continuous interval $[\alpha,\alpha+\Delta]$ for growing $\Delta$ over several \emph{phases} (inspect Figure~\ref{fig:theorem7}). In the $l$th phase, we use a constant number of $2^l$-approximate queries.
\begin{itemize}
    \item If at any point we cover $[\alpha,\alpha+\Delta]$ such that $\Delta\geq (\beta-\alpha)/2$, we stop.
        \item If we never cover past $\alpha+\Delta$ with $\Delta\geq (\beta-\alpha)/2$ in phase $l$, we do exactly three $2^l$-approximate queries in that phase:
        \item In phase 0, we do regular \SSI queries for $\alpha$, $\alpha+1$, $\alpha+2$. \item In phase $l$, if we have covered up until $\alpha+\Delta$ in phase $l-1$, we center the first $2^l$-approximate query at the largest $\kappa\cdot 2^l$ such that $\kappa\cdot 2^l-2^{l-1}\leq \alpha+\Delta$. Then we center the next two queries at $(\kappa+1)\cdot 2^l$ and $(\kappa+2)\cdot 2^l$.
        \item When we stop, we do a symmetric process starting from $\beta$.
        \end{itemize}
        \begin{figure}
            \centering
            \includegraphics[width=0.82\textwidth]{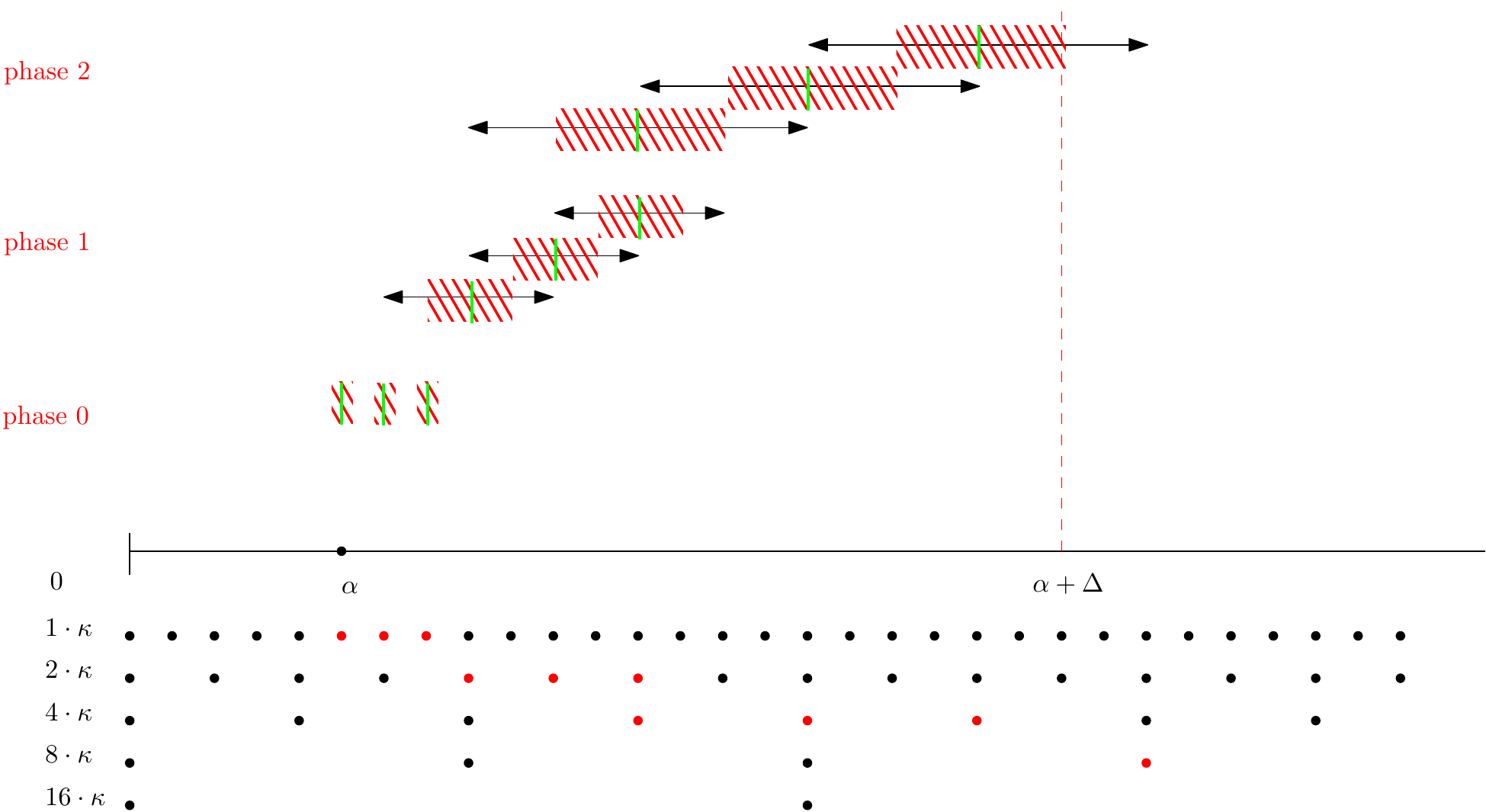}
            \caption{Illustration of the phases from the proof of Theorem~\ref{thm:intervals}: At the $l$th phase of the algorithm, we use at most three $2^{l}$-approximate queries to cover at least $2\cdot 2^l$ elements in $[\alpha,\beta]$ which were not covered by previous phases. The centers of the queries are shown by short vertical line segments. The covered elements for each query are shown dashed, and the uncertain elements are shown by the black horizontal arrows. The dots at the bottom show all potential centers for the approximate queries; the ones our algorithm uses are marked in red.}
            \label{fig:theorem7}
        \end{figure}
        \begin{claim}\label{claim:cover}
        We stop after $\cO(\log (\beta-\alpha))$ phases, after which we will have covered the full interval $[\alpha,\alpha+\Delta]$ for $\Delta\geq (\beta-\alpha)/2$.
        \end{claim}
        \begin{claimproof} By the choice of the first query in each phase, the interval we cover with that query starts at some position $\kappa\cdot 2^l-2^{l-1}\leq \alpha+\Delta$. Thus, there are no gaps. The first interval covers up until $\kappa\cdot 2^l+2^{l-1}=(\kappa+1)\cdot 2^l-2^{l-1}>\alpha+\Delta$. Thus, only the first query of a phase covers part of an interval that was covered in a previous phase, while the next two queries cover 
        $2\cdot 2^l$ elements which were not covered in a previous phase. Thus, after $\cO(\log(\beta-\alpha))$ phases, we have covered at least until $\alpha+(\beta-\alpha)/2$.
        \end{claimproof}
        \begin{claim}\label{claim:uncertain}
        All uncertain elements introduced during the algorithm are included in $[\alpha,\beta]$.
        \end{claim}
        \begin{claimproof}
 By the argument before, if we enter phase $l$ and have not stopped, we have covered at least $2\sum_{j=0}^{l-1} 2^j=2(2^l-1)=2^{l+1}-2$ elements, i.e.,  $\Delta\geq 2^{l+1}-2$. Further, $\Delta<(\beta-\alpha)/2$. The first query in phase $l$ is centered at $\kappa\cdot 2^l$ with $\kappa\cdot 2^l-2^{l-1}\leq \alpha+\Delta < \kappa\cdot 2^l+2^{l-1}$. Thus, $\kappa\cdot 2^l\leq \alpha+\Delta+2^{l-1}$ and $\kappa\cdot 2^l\geq\alpha+\Delta-2^{l-1}+1$ and the uncertain interval $(\kappa\cdot 2^l-2^l,\kappa\cdot 2^l+2^l)$ is contained in  $[\alpha+\Delta-2^{l-1}+2-2^l,~\alpha+\Delta+2^{l-1}+2^l-1]$.
Since $\Delta\geq 2^{l+1}-2\geq 2^l+2^{l-1}-2$, we have that $\alpha+\Delta-2^{l-1}-2^l+2\geq \alpha$ and thus there are no uncertain elements smaller than $\alpha$.
Since $\Delta+2^l+2^{l-1}-1\leq 2\Delta+1\leq \beta-\alpha$, we have that $\alpha+\Delta+2^l+2^{l-1}-1\leq \beta$ and thus there are no uncertain elements bigger than $\beta$.

If there are subsequent queries in phase $l$, then before the query we cover up to $\alpha+\Delta$ for $\Delta<(\beta-\alpha)/2$. The next query is centered at $\kappa\cdot2^l=\alpha+\Delta+2^{l-1}$, thus the argument as to why we do not introduce uncertain elements past $\beta$ is analogous. Since $\Delta\geq 2^l$ and $\kappa\cdot 2^l>\alpha+\Delta$, the uncertain interval centered at $\kappa\cdot 2^l$ cannot include elements smaller than $\alpha$.
\end{claimproof}

Combining Claim~\ref{claim:cover} and Claim~\ref{claim:uncertain}, we obtain the theorem.
    \end{proof}

Next, we show how we can solve \SSIintervalsrep, the reporting variant of the \SSIintervals\ problem. 
The reduction is basically the same, but using a solution to \SSIreport\ instead of a solution to \SSI. Again, we use an approximate variant of the problem, now defined as follows:

\defDSproblem{\SSIintervalsapproxrep}{A collection of $k$ sets $S_1, \ldots, S_k$ of total size $\sum_i |S_i| = N$ of integers from a universe $U = \{1,2, \dots, u\}$ and a \emph{level} $l$ with $1\leq l\leq \log u$.}{Given $i,j$ and a \emph{center distance} $d=\kappa\cdot 2^l$, for some positive integer $\kappa$, output a set $P$ of pairs $(a,b)$ such that $a\in S_i$ and $b\in S_j$ and
\begin{itemize}
    \item $P$ contains all pairs $(a,b)$, $a\in S_i$, $b\in S_j$ such that $b-a\in[d-2^{l-1},d+2^{l-1}]$;
    \item $P$ contains no pair $(a,b)$ such that $b-a\notin(d-2^l,d+2^l)$.
\end{itemize}}

    \begin{corollary}\label{cor:approxrep}
        For every $0\leq \delta \leq 1$, there is a data structure for \SSIintervalsapproxrep with $\ctO(N^{2-\delta/3})$ space and $\ctO(N^{\delta}\cdot |P|)$ query time.
    \end{corollary}
\begin{proof}
The reduction is essentially the same as Lemma~\ref{lem:approxinterval}, just that instead of the \SSI data structure, we use the \SSIreport\ data structure from Theorem~\ref{thm:reporting}. That is, we construct a \SSIreport data structure for sets $S'_i=\{\lfloor a/2^{l-1}\rfloor, a\in S_i\}$. Additionally, for any $a'\in S'_i$, we store a list $L_{a'}$ of all values $a\in S_i$ such that $\lfloor a/2^{l-1}\rfloor=a'$. To answer a query $(i,j,d=\kappa\cdot 2^l)$, we query the \SSIreport\ data structure for $(i,j,2\kappa-1),(i,j,2\kappa)$ and $(i,j,2\kappa+1)$. Whenever one of the queries returns a pair $(a',b')$, we return \emph{all} pairs $(a,b)$ for $a\in L_{a'}$ and $b\in L_{b'}$. By the same arguments as in the proof of Lemma~\ref{lem:approxinterval}, we never return a pair with $b-a\notin(d-2^l,d+2^l)$, and we return all pairs $b-a\in[d-2^{l-1},d+2^{l-1}]$.
\end{proof}

\intervalsrep*

\begin{proof}
        We use the same reduction as in Theorem~\ref{thm:intervals}, only using the data structures for \SSIintervalsapproxrep from Corollary~\ref{cor:approxrep} instead of the data structures for \SSIintervalsapprox. We perform exactly the same queries. By the same arguments as in the proof of Theorem~\ref{thm:intervals}, we find all pairs $(a,b)$ with $a\in S_i$ and $b\in S_j$ and $b-a\in[\alpha,\beta]$ and only those. However, we might report the same pair many times, so we need to argue about the total size of the output. Note that any single query to \SSIintervalsapproxrep reports any pair at most once, thus, any pair is reported $\cO(\log u)$ times and the total size of the output is $\cO(\occ\cdot \log u)$. To avoid multiple outputs we can collect all pairs, sort them, and delete duplicates, before outputting. 
\end{proof}

\section{\GI}\label{sec:GI}

Let us recall some basic definitions and notation on strings.
An \emph{alphabet} $\Sigma$ is a finite set of elements called \emph{letters}. A \emph{string} $S=S[1\dd n]$ is a sequence of letters over some alphabet $\Sigma$; we denote the length of $S$ by $|S|=n$. The fragment $S[i\dd j]$ of $S$ is an \emph{occurrence} of the underlying \emph{substring} $P=S[i]\ldots S[j]$. We also write that $P$ occurs at \emph{position} $i$ in $S$ when $P=S[i] \ldots S[j]$. A {\em prefix} of $S$ is a fragment of $S$ of the form $S[1\dd j]$ and a {\em suffix} of $S$ is a fragment of $S$ of the form $S[i\dd n]$.

For a string $S$ of length $n$ over an ordered alphabet of size $\sigma$, the \emph{suffix array} $\textsf{SA}[1\dd n]$ stores the permutation of $\{1,\ldots, n\}$ such that $\textsf{SA}[i]$ is the starting position of the $i$th lexicographically smallest suffix of $S$. The standard $\textsf{SA}$ application is as a text index, in which $S$ is the \emph{text}: given any string $P[1\dd m]$, known as the \emph{pattern}, the suffix array of $S$ allows us to report all $\occ$ occurrences of $P$ in $S$ using only $\cO(m \log n + \occ)$ operations~\cite{DBLP:journals/siamcomp/ManberM93}. 
We do a binary search in $\textsf{SA}$, which results in an interval $[s,e)$ of suffixes of $S$ having $P$ as a prefix. Then, $\textsf{SA}[s\dd e-1]$ contains the starting positions of all occurrences of $P$ in $S$. The $\textsf{SA}$ is often augmented with the $\textsf{LCP}$ array~\cite{DBLP:journals/siamcomp/ManberM93} storing the length of longest common prefixes of lexicographically adjacent suffixes. In this case, reporting all $\occ$ occurrences of $P$ in $S$ can be done in $\cO(m+ \log n + \occ)$ time~\cite{DBLP:journals/siamcomp/ManberM93} (see~\cite{DBLP:journals/algorithmica/0001KL15,DBLP:conf/cpm/0001G15,DBLP:journals/siamcomp/NavarroN17} for subsequent improvements). The suffix array can be constructed in $\cO(n)$ time for an integer alphabet of size $\sigma=n^{\cO(1)}$~\cite{DBLP:conf/focs/Farach97}.
Given the suffix array of $S$, we can compute the \textsf{LCP} array of $S$ in $\cO(n)$ time~\cite{DBLP:conf/cpm/KasaiLAAP01}. 

The \emph{suffix tree} of $S$, which we denote by $\textsf{ST}(S)$, is the compacted trie of all the suffixes of $S$~\cite{DBLP:conf/focs/Weiner73}. Assuming $S$ ends with a unique terminating symbol, every suffix $S[i\dd n]$ of $S$ is represented by a leaf node that we decorate with $i$. We refer to the set of indices stored at the leaf nodes in the subtree rooted at node $v$ as the \emph{leaf-list} of $v$, and we denote it by $LL(v)$. Each edge in $\textsf{ST}(S)$ is labeled with a substring of $S$ such that the path from the root to the leaf annotated with index $i$ spells the suffix $S[i\dd n]$. We refer to the substring of $S$ spelled by the path from the root to node $v$ as the \emph{path-label} of $v$ and denote it by $L(v)$. Given any pattern $P[1\dd m]$, the suffix tree of $S$ allows us to report all $\occ$ occurrences of $P$ in $S$ using only $\cO(m \log \sigma + \occ)$ operations.
We spell $P$ from the root of $\textsf{ST}(S)$ (to access edges by the first letter of their label, we use binary search) until we arrive (if possible) at the first node $v$ such that $P$ is a prefix of $L(v)$. Then all $\occ$ occurrences of $P$ in $S$ are precisely $LL(v)$. The suffix tree can be constructed in $\cO(n)$ time for an integer alphabet of size $\sigma=n^{\cO(1)}$~\cite{DBLP:conf/focs/Farach97}.
To improve the query time to $\cO(m + \occ)$ we use randomization to construct a perfect hash table~\cite{FKS1984} accessing edges by the first letter of their label in $\cO(1)$ time.

We next show how to reduce \GIrep to \SSIintervalsrep. A query for $P_1$, $P_2$ and $[\alpha,\beta]$ corresponds to a \SSIintervalsrep query for $S_1$, $S_2$ and $[\alpha,\beta]$, where $S_1$ is the set of occurrences of $P_1$ in $S$ and $S_2$ is the set of occurrences of $P_2$ in $S$. An obvious strategy would be to preprocess all sets which correspond to leaves below a node in the suffix tree into a \SSIintervalsrep data structure. The issue is that the total size of these sets can be $\Omega(n^2)$. However, any such set corresponds to a consecutive interval within the suffix array. Thus, the strategy is as follows: We store the suffix tree and the suffix array for $S$. We cover the suffix array in dyadic intervals and preprocess the resulting subarrays into the \SSIintervalsrep data structure. 
In detail, let $D$ be the set of dyadic intervals covering $[1,n]$. That is, $D$ includes all intervals of the form $[1+\kappa\cdot 2^j, (\kappa+1)\cdot 2^j]$ for all $0\leq \kappa\leq \lfloor \frac{n}{2^j}\rfloor -1$ and $0\leq j\leq \lfloor \log n \rfloor$. For any interval $[\gamma_1,\gamma_2]\in D$, we define a set containing the elements in $\textsf{SA}[\gamma_1\dd \gamma_2]$. We preprocess all of these sets into the \SSIintervalsrep data structure.
The total size of the sets in  the data structure is $\cO(n\log n)$. For a query, we find the suffix array intervals $(I_1,I_2)$ for $P_1$ and $P_2$, respectively, in $\cO(|P_1|+|P_2|)$ time, using standard pattern matching in the suffix tree. Now, let $\mathcal{A}$ be the collection of sets corresponding to dyadic intervals covering $I_1$ and $\mathcal{B}$ the collection of sets corresponding to dyadic intervals covering $I_2$. We can find all pairs $(i,j)$ of occurrences of $P_1$ and $P_2$ satisfying $j-i\in [\alpha,\beta]$ by querying \SSIintervalsrep for $(A,B,[\alpha,\beta])$ for all $A\in \mathcal{A}$ and $B\in \mathcal{B}$. 

In conclusion, we have shown the following:

\begin{theorem}\label{thm:SSItoGI}
    Assume there is a data structure for \SSI with $s(N)$ space and $t(N)$ query time, where $N$ is the input size, and which outputs a witness pair $(a,b)\in S_i\times S_j$ satisfying $a+s=b$ for a query $(i,j,s)$. Then there is a data structure for \GIrep with $\ctO(n+s(n))$ space and $\ctO( |P_1| + |P_2| + t(n)\cdot(\occ + 1))$ query time, where $n$ is the length of the input string and $\occ$ is the size of the output.
\end{theorem}

The following two corollaries follow from Theorem~\ref{thm:SSItoGI} together with Theorem~\ref{thm:intervalsrep} and Lemma~\ref{lem:diftrade-off}, respectively.

\begin{corollary}
    For every $0\leq \delta \leq 1$, there is a data structure for \GIrep with $\ctO(n^{2-\delta/3})$ space and $\ctO(|P_1|+|P_2|+n^{\delta}\cdot (\occ+1))$ query time, where $\occ$ is the size of the output.
\end{corollary}

\begin{corollary}
    For every $0\leq \delta \leq 1$, there is a data structure for \GIrep with $\ctO(n^{3-2\delta})$ space and $\ctO(|P_1|+|P_2|+n^{\delta}\cdot (\occ+1))$ query time, where $\occ$ is the size of the output.
\end{corollary}
Note that $n^{3-2\delta}$ is smaller than $n^{2-\delta/3}$ for $\delta>3/5$.
We have arrived at Theorem~\ref{the:main}.

\section{\JI}\label{sec:JI}
A solution to \THREESUMDSreport implies solutions to other problems. As proof of concept, we show here the implications to \JIrep: preprocess a string $S$ of length $n$ over an alphabet $\Sigma$ into a compact data structure that facilitates efficient search for substrings of $S$ whose characters have a specified histogram.
For instance, $P=acaacabd$ has the histogram $h(P)=(4,1,2,1)$: $h(P)[a]=4$ because $a$ occurs $4$ times in $P$, $h(P)[b]=1$, etc. The existence variant, \JI, answers the question of whether such a substring occurs in $S$ or not. We show the following result:

\corjumbled*

Before proving the above statement, we will briefly overview the related literature.
    
\subparagraph{Related work.}~Previous work on \JI has achieved the bounds of 
 $\cO(n^{2-\delta})$ space and $\cO(m^{(2\sigma-1)\delta})$ query time~\cite{journals/algorithmica/KociumakaRR17}, where $m$ is the norm of the pattern and $\sigma=|\Sigma|$.
 Later, a data structure with
 $\ctO(n^{2-\delta})$ space and $\ctO(n^{\delta(\sigma+1)/2})$ query time was presented~\cite{conf/stoc/ChanL15}. 

For the special case of a binary alphabet, more efficient algorithms exist; namely, $\cO(n)$ space and $\cO(1)$ query time~\cite{conf/stringology/CicaleseFL09}. The data structure in~\cite{conf/stringology/CicaleseFL09} has a preprocessing time of $\cO(n^2)$, which can be improved to $\ctO(n^{1.5})$ using the connection to min-plus-convolution~\cite{conf/stoc/ChanL15,conf/stoc/ChiDX022}.

\JI has also seen interest from the lower-bound side, where~\cite{conf/soda/AfshaniDKN20} shows that if a data structure can report all the $\occ$ matches to a histogram query in $\cO(n^{0.5-o(1)} + \occ)$ time, then it needs to use $\Omega(n^{2-o(1)})$ space, even for the special case of a binary alphabet. 
In fine-grained complexity, the problem has also received attention;~\cite{conf/icalp/AmirCLL} shows that under a 3SUM hardness assumption, for alphabets of size $\omega(1)$, we cannot get $\cO(n^{2-\epsilon})$ preprocessing time and $\cO(n^{1-\delta})$ query time for any $\epsilon,\delta>0$. Furthermore, under the now refuted Strong \THREESUMDS conjecture,~\cite{DBLP:conf/wads/GoldsteinKLP17} gives conditional lower bounds, that are contradicted by our results. In particular, their conditional lower bound (conditioned on the now refuted Strong \THREESUMDS conjecture) argues against a $\cO(n^{2-\frac{2(1-\alpha)}{\sigma-1-\alpha}-\Omega(1)})$ space and $\cO(n^{1-\frac{1+\alpha(\sigma-3)}{\sigma-1-\alpha}-\Omega(1)})$ query time solution for any $0\leq \alpha\leq 1$. Setting $\alpha=0$ and $\sigma=9$, this would imply that there does not exist a $\cO(n^{2-\frac{1}{4}-\Omega(1)})$ space and $\cO(n^{1-\frac{1}{8}-\Omega(1)})$ query time solution. However, setting $\delta=1/4+\epsilon$ for $\epsilon=1/48$, we obtain an $\ctO(n^{2-\frac{1}{4}-\epsilon})$ space and $\ctO(n^{\frac{3}{4}+3\epsilon})=\ctO(n^{\frac{6}{8}+\frac{1}{16}})=\ctO(n^{1-\frac{1}{8}-\frac{1}{16}})$ query time, a contradiction. We now return back to the proof of Corollary~\ref{cor:jumbled}:

\begin{proof}[Proof of \cref{cor:jumbled}] 
Let us first remind the reader of the well-known reduction to $d$-dimensional \THREESUMDS, as introduced by Chan and Lewenstein in~\cite{conf/stoc/ChanL15}. The $d$-dimensional \THREESUMDS problem is defined as follows: Preprocess two sets $A$ and $B$ of $N$ vectors from $[0,\dots,u-1]^d$ each such that we can efficiently answer queries of the following form: for some $c\in[0,\dots,u-1]^d$, decide if there exists $a\in A$ and $b\in B$ such that $c=a+b$.

We reduce \JI to $\sigma$-dimensional \THREESUMDS as follows. Let $\sigma=|\Sigma|$. 
Let $A$ be the set of all histograms of prefixes of the input string $S$, 
and similarly, let $B$ be the set of all histograms of suffixes of $S$. 
We thus have that the cardinality of these two sets is the length of the string $S$, i.e., $|A|=|B|=|S|=n$, and may thus build a $\sigma$-dimensional \THREESUMDS data structure for $A$ and $B$.
Additionally, we store the histogram $h(S)$ of $S$. 
For a query histogram $P$, compute $c=h(S)-P$, and query the $\sigma$-dimensional \THREESUMDS data structure for $c$. Any match $a,b$ returned by the data structure corresponds to a histogram of the complement of some substring $p$ whose histogram is $P$: $a$ corresponds to the prefix of everything before $p$, and $b$ corresponds to the suffix of everything after $p$. 

Now, we note that there is a reduction from $d$-dimensional \THREESUMDS over a universe of size $u$ to \THREESUMDS over a universe of size $\cO(u^d)$. Our reduction works by doing the following transformation on sets $A$ and $B$: define $A'=\{a_1+a_2\cdot u + a_3\cdot u^2 + \dots a_d\cdot u^{d-1}: a\in A\}$; and define $B'$ in the same way. For a query $c$, define the query $c'=c_1+c_2\cdot u + c_3\cdot u^2\dots c_d\cdot u^{d-1}$. Thus, by spacing out using $u$-factors, we ensure that any match to a query $c$ to the sets $A$ and $B$ corresponds exactly to a match to the corresponding query $c'$ to the sets $A'$ and $B'$.
With this reduction, the set size remains unchanged, that is, $|A|=|A'|$ and $|B|=|B'|$, however, we are now indexing with a universe of size $u'=O(u^d)$.

Thus, finally, for \JI, let $n$ be the length of the string $S$. Our reductions would result in a universe of size $u'=n^{\sigma}$. For constant $\sigma$ this value $u'$ is polynomial in $n$. Therefore, by Corollary~\ref{thm:3sum-reporting}, we obtain a data structure for \JI over constant sized alphabets 
with $\ctO(n^{2-\delta})$ space and $\ctO(n^{3\delta}\cdot(\occ+1))$ query time.
\end{proof}

\section{Final Remarks}\label{sec:con}

Our solutions show new and interesting relations between \GI, \SSI, and \THREESUMDS; in particular, we contribute new trade-offs for \THREESUMDSreport. Chan showed that the \THREESUM problem has direct applications in computational geometry~\cite{DBLP:journals/talg/Chan20};
it would be interesting to see if our data structure yields improved bounds for the data structure versions of these geometric problems.

\bibliography{references}
\vfill
\pagebreak
\appendix

\section{A Linear-Space Solution for Gapped Indexing}\label{sec:linear_space}

The linear-space solution uses a standard linear-time pattern matching algorithm (e.g., the Knuth-Morris-Pratt algorithm~\cite{DBLP:journals/siamcomp/KnuthMP77}), which, given two strings $S$ and $P$, it outputs all positions in $S$ where $P$ occurs, in sorted order. We run this algorithm two times: on $P_1$ and $S$; and on $P_2$ and $S$. The two sorted lists of positions are scanned from left to right using a two-finger approach: We start with the first position $i_1$ where $P_1$ occurs in $S$, and scan the list of $P_2$ occurrences until we find the first one that is at least $i_1+\alpha$. Then we scan and output all pairs until we arrive at a position of $P_2$ which is larger than $i_1+\beta$. Then we advance to the next position $i_2$ of $P_1$, and from the last position we considered in the list for $P_2$, we scan first backward and then forward to output all positions in $[i_2+\alpha,i_2+\beta]$. Every position we scan is charged to an occurrence we output, thus we achieve a running time of $\cO(n+\occ)$.

\section{A Near-Quadratic-Space Solution for Gapped Indexing}\label{sec:quadratic_space}

Let us start with a data structure of $\cO(n^3)$ space and $\cO(|P_1| + |P_2| + \occ)$ query time: We store, for each pair $(v, w)$ of nodes in the suffix tree of $S$, the set of distances that appear between the set of occurrences defined by $v$ and $w$. To answer a query, we search for $P_1$ and $P_2$ in the suffix tree to locate the corresponding pair of nodes and then report the distances within the gap range $[\alpha, \beta]$. Since the number of distinct distances is at most $n-1$ this leads to a solution using $\cO(n^3)$ space and $\cO(|P_1| + |P_2| + \occ)$ query time, where $\occ$ is the size of the output. A more sophisticated approach can reduce the space to $\ctO(n^2)$ at the cost of increasing the query time by polylogarithmic factors. Namely, the idea is to consider subarrays of the suffix array of $S$~\cite{DBLP:journals/siamcomp/ManberM93} corresponding to the dyadic intervals on $[1,n]$, and store, for every possible distance $d$, all pairs of subarrays $(A,B)$ containing elements at distance $d$. Further, we store a list of all $(a,b)$, $a\in A$ and $b\in B$ such that $b-a=d$. Since the total size of the subarrays is $\cO(n\log n)$, the total number of pairs of elements is $\cO(n^2\log^2 n)$, thus this solution can be stored using $\ctO(n^2)$ space. We also store the suffix tree of $S$, which takes $\Theta(n)$ extra space. To answer a query, we search for $P_1$ and $P_2$ in the suffix tree to find their suffix array intervals: their starting positions in $S$ sorted lexicographically with respect to the corresponding suffixes. We cover each interval in dyadic intervals and use the precomputed solution for any pair consisting of a subarray in the cover of $P_1$'s interval and a subarray in the cover of $P_2$'s interval. This results in query time $\cO(|P_1|+|P_2|+\log^2 n\cdot(\occ+1))$.

\section{Smallest Shift}\label{sec:smallest_shift}

When studying the problem of deciding whether a given shift incurs an intersection between sets, a natural next question is the optimization (minimization) variant of the problem: \emph{what is the smallest shift that yields an intersection?} We formally define this problem next.

\defDSproblem{Smallest Shift}{A collection of $k$ sets $S_1, \ldots, S_k$ of total size $\sum_i |S_i| = N$ of integers from a universe $U = \{1,2, \ldots, u\}$.}{Given $i,j$, output  
the smallest $s$ such that there exists $a\in S_i$ and $b\in S_j$ with $a+s=b$.}

Note that the problem is symmetric in $i$ and $j$ in the sense that finding the smallest positive shift transferring $S_i$ to intersect $S_j$ is equivalent to finding the largest negative shift transferring $S_j$ to intersect $S_i$. Note also that by constructing a predecessor/successor data structure~\cite{WILLARD198381} at preprocessing time, we may answer the query in $\ctO(|S_i|)$ time: simply perform a successor (or predecessor) query in $S_j$ for each element $a\in S_i$, and report the smallest element-successor distance. 

Thus, we are safely able to handle \SSHIFT queries in $\cO(\sqrt{N}\log\log N )$ time in all cases where either $S_i$ or, because of symmetry, $S_j$, is of size at most $\sqrt{N}$. Remaining is the case where both sets are of size larger than $\sqrt{N}$.  Here, however, we note that at most $\cO(\sqrt{N})$ such sets can exist, so we can tabulate all answers in $\cO(\sqrt{N}\cdot \sqrt{N}) = O(N)$ space. 

In other words, we have the following solution:

\noindent\subparagraph{Preprocessing:} 
Store for each $i$ whether $|S_i|\le \sqrt{N}$. Let $l$ denote the number of sets $L_1,\ldots,L_l$ of size larger than $\sqrt{N}$. For each set $S_i$ with $|S_i|>\sqrt{N}$ store its index in the list of large sets $L_1,\ldots, L_l$. \begin{itemize}[noitemsep, nolistsep]
\item For all $i$, with $|S_i|\le \sqrt{N}$,
compute and store predecessor $\operatorname{pre}(S_i)$ and successor $\operatorname{suc}(S_i)$ data structures for the set $S_i$. \item Store an $l\times l$ table in which the $(i,j)$th entry stores the smallest shift $s$ such that there exists $a\in L_i$ and $b\in L_j$ with $a+s=b$.
\end{itemize}

\subparagraph{Query $i,j$:}
\begin{itemize}[noitemsep, nolistsep]
\item If $|S_i|\le \sqrt{N}$, perform $|S_i|$ successor queries in $\operatorname{suc}(S_j)$, and return the smallest difference.%
\item If $|S_j|\le \sqrt{N}$, similarly, perform $|S_i|$ predecessor queries in $\operatorname{pre}(S_i)$.%
\item If $|S_i|> \sqrt{N}$ and $|S_j|> \sqrt{N}$, use the precomputed information to output the smallest shift.
\end{itemize}

\begin{proposition}
The above data structure for \SSHIFT uses $\cO(N)$ space and has $\ctO(\sqrt{N})$ query time. Moreover, the data structure can be constructed in $\cO(N \sqrt{N})$ time.
\end{proposition}

\begin{proof}
We construct predecessor and successor queries for all sets in $\sum_i |S_i|\log\log |S_i| = O(N\log \log N)$ total time and $\sum_i |S_i| = O(N)$ space, with $\cO(\log \log N)$ query time~\cite{WILLARD198381}, surmounting to an $\cO(\sqrt{N}\log\log N)$ query time when either $S_i$ or $S_j$ is smaller than $\sqrt{N}$.

For constructing the $l\times l$ table,  we note that $l\le\sqrt{N}$. Since the $(i,j)$th entry of the table can be computed in time $\cO(|L_i| + |L_j|)$ by a merge-like traversal of the sorted respective sets, the total preprocessing time becomes:

\begin{equation*} 
\begin{split}
\sum_i \sum_{j\neq i} (|L_i| + |L_j|) \le \sum_i (\sqrt{N}|L_i| +\sum_j |L_j|) \le \sum_i ( \sqrt{N} |L_i| + N) \\
= \sqrt{N}\sum_i |L_i|+\sum_i N = \cO(\sqrt{N}N).
\end{split}
\end{equation*}

Here, we are using that $\sum_i |S_i| = N$ and that there are $l\le\sqrt{N}$ large sets. This $\cO(N\sqrt{N})$ term dominates the $\sum_i \ctO(|S_i|)$ terms needed to sort the sets and construct predecessor/successor data structures for each of them. 
\end{proof}

\end{document}